\documentclass[sigconf, nonacm]{acmart}
\usepackage{amsmath}
\usepackage{amsthm}
\usepackage{algorithm}
\usepackage[noend]{algpseudocode}
\usepackage{balance}
\usepackage{booktabs}
\usepackage[font=small,skip=1pt]{caption}
\usepackage{color,colortbl}
\usepackage{comment}
\usepackage{float}
\usepackage{graphicx}
\usepackage[first=0,last=9]{lcg}
\usepackage{mathtools}
\usepackage{multicol}
\usepackage{paralist}
\usepackage{subcaption}
\usepackage{wrapfig}
\usepackage{xspace}
\usepackage[capitalise]{cleveref}
\usepackage[subtle]{savetrees}

\newcommand{\namenospace}{\textsc{Odin}}
\newcommand{\name}{\textsc{Odin} }
\usepackage{enumitem}
\setlist{topsep=0pt, leftmargin=*}

\begin{document}

\title{\namenospace: A NL2SQL Recommender to Handle Schema Ambiguity}

\author[K. Vaidya, A. Sankararaman, J. Ding, C. Lei, X. Qin, B. Narayanaswamy, T. Kraska]
{
    Kapil Vaidya$^{1*}$, Abishek~Sankararaman$^2$, Jialin Ding$^2$, Chuan Lei$^2$, Xiao Qin$^2$, Balakrishnan Narayanaswamy$^2$, Tim Kraska$^2$
}
\affiliation{%
  \institution{$^1$Parallel Web Systems\quad$^2$Amazon Web Services}
  \city{}
  \country{}
}
\thanks{$^*$Work performed while employed at AWS}

\begin{abstract}
NL2SQL (natural language to SQL) systems translate natural language into SQL queries, allowing users with no technical background to interact with databases and create tools like reports or visualizations. While recent advancements in large language models (LLMs) have significantly improved NL2SQL accuracy, schema ambiguity remains a major challenge in enterprise environments with complex schemas, where multiple tables and columns with semantically similar names often co-exist. To address schema ambiguity, we introduce \namenospace, a NL2SQL recommendation engine. Instead of producing a single SQL query given a natural language question, \name generates a set of potential SQL queries by accounting for different interpretations of ambiguous schema components. \name dynamically adjusts the number of suggestions based on the level of ambiguity, and \name learns from user feedback to personalize future SQL query recommendations. Our evaluation shows that \name improves the likelihood of generating the correct SQL query by 1.5--2$\times$ compared to baselines.
\end{abstract}

\maketitle

\section{Introduction}
\label{sec:intro}

NL2SQL (natural language to SQL) systems translate natural language questions into SQL queries, allowing users with no technical background to interact with databases and create tools like reports or visualizations. For example, a manager could ask, “\textit{How many new customers did we acquire this quarter?},” and the system would generate the corresponding SQL query, execute it, and return the results in an accessible format. This broadens access to data-driven insights, making NL2SQL an essential tool for decision-making. Large language models (LLMs) have played a transformative role in advancing NL2SQL technology. Their ability to adapt to unfamiliar tasks by leveraging contextual examples allows them to generate accurate outputs. This adaptability has positioned LLM-based solutions at the top of various well-known NL2SQL benchmarks like Spider and BIRD~\cite{yu2019spider,li2023llmBird}.

In enterprise environments, database schemas are often large and highly complex due to the integration of multiple data sources, frequent schema versioning, and table transformations. This complexity introduces ambiguity when translating natural language questions into SQL queries. Specifically, tables and columns with similar names can exist, making it difficult to identify the user's intended reference. We refer to this challenge as \textit{schema ambiguity}, where multiple SQL queries, each using different tables or columns, could potentially be correct.

For instance, a user might ask, ``\textit{What are the average salaries by department?}'', but the database schema could contain both a \texttt{curr\_dept} table and a \texttt{dept\_2022} table. In this scenario, it is unclear which table the user intends to query, leading to multiple potential SQL queries. Similarly, ambiguity can arise when dealing with column names. For example, when a user asks, ``\textit{What were the total sales last quarter?}'', the \texttt{sales} table may have columns named both \texttt{gross\_sales} and \texttt{net\_sales}, either of which could be valid depending on the user's intent. 

Furthermore, schema ambiguity can substantially impact the structure of a SQL query. For example, consider the question, ``\textit{What is the revenue per customer?}''. In this case, the database schema might include both a \texttt{customers} table and an \texttt{orders} table, where the \texttt{orders} table contains a column named \texttt{revenue}, while the \texttt{customers} table may include an aggregated column, \texttt{total\_revenue}. Depending on which table the user intends to reference, two structurally different SQL queries can be generated:

\begin{verbatim}
SQL1: SELECT customers.customer_id, SUM(orders.revenue)
FROM customers
JOIN orders ON customers.customer_id = orders.customer_id
GROUP BY customers.customer_id;
\end{verbatim}

\begin{verbatim}
SQL2: SELECT customer_id, total_revenue
FROM customers;
\end{verbatim}

SQL1 involves a join and aggregation across multiple rows in the \texttt{orders} table, while SQL2 simply retrieves pre-aggregated data from the \texttt{customers} table. This example demonstrates that schema ambiguity can have an effect on the complexity and runtime performance of the SQL query, making its resolution a critical aspect of NL2SQL systems.

Among multiple potential SQL queries, users often prefer one query over the others due to their preferences toward specific tables or columns in the schema.
These preferences may reflect the user’s familiarity with certain data sources or their expectations regarding the relevance of certain schema components. For instance, the user may prefer \textit{gross\_sales} instead \textit{net\_sales} for revenue analysis due to the business logic of their organization. By learning these preferences, a NL2SQL system could provide more personalized query suggestions, enhancing the user experience.

One approach to address schema ambiguity is to generate a set of SQL queries that accounts for all possible interpretations of the ambiguous schema components. This allows users to review the options and select the query that best fits their needs. However, generating such a set is challenging. LLM-based NL2SQL systems often struggle to produce a sufficiently diverse set of potential SQL queries, even when using techniques such as sampling methods or temperature adjustments, as highlighted by~\cite{ambiqt}. Additionally, the set cannot be too large, as it would overwhelm the user and make selecting the correct query difficult. 
The complexity increases further when users have specific preferences for tables or columns which need to be captured and incorporated into future questions.


To address the challenges presented by schema ambiguity, we introduce \namenospace, a NL2SQL recommendation engine that helps users manage and resolve schema ambiguity in NL2SQL tasks. Instead of returning a single SQL query, \name presents a set of potential queries for users to choose from. The number of suggestions is dynamically adjusted based on the level of ambiguity in the user’s question. Additionally, \name can learn user's preferences for specific schema components through their feedback, enhancing future recommendations.

Internally, \name operates using a Generate-Select paradigm. Given a semantically ambiguous natural language question, the \textbf{Generator} is tasked with producing the set of all potentially correct SQL queries. Traditional approaches, such as beam search or diversity-promoting techniques like nucleus sampling~\cite{holtzman2019curious}, often fail to generate queries that reflect schema ambiguity, as noted in~\cite{ambiqt}. To overcome this limitation, \name uses an iterative strategy: the Generator sequentially produces candidate SQL queries by selectively modifying the information provided to the LLM. Specifically, it removes certain schema elements from previously generated SQL queries, encouraging the LLM to explore different schema components and generate diverse queries.

The generator may produce incorrect SQL queries by omitting too many important schema components or by generating results that do not align with user preferences. The \textbf{Selector} component addresses this by filtering out the inaccuracies to reduce the set size while ensuring the correct SQL query is retained, thus maintaining high recall. This process is framed within the conformal prediction framework~\cite{conformaltutotrial}, which is commonly used to provide confidence in the outputs of machine learning models in critical fields such as medical diagnosis~\cite{vazquez2022conformal}. Conformal prediction provides concrete guarantees on recall, ensuring that it does not fall below a specified threshold.

After the selection phase, users can choose their preferred SQL query from the remaining options. \name learns user preferences from this feedback, allowing it to refine its recommendations and better align future outputs with the user’s specific preferences.

Our evaluation demonstrates that the set of SQL queries recommended by \name contains the correct query 1.5--2$\times$ more often, while maintaining a result set that is 2--2.5$\times$ smaller, compared to other baselines on the AmbiQT benchmark, which contains various forms of schema ambiguity\footnote{We use a modified version of this benchmark to introduce additional ambiguity in the questions.}.

In summary, we make the following contributions: 

\begin{itemize}
    \item We present \namenospace, a system designed to handle and resolve ambiguity in NL2SQL tasks (Section \ref{sec:odin_overview}).
    \item We introduce a novel SQL generation-selection strategy that outperforms LLM-based sampling (Sections \ref{sec:generation} and \ref{sec:selection}).
    \item We offer a personalization algorithm for \name to learn user preferences (Section \ref{sec:personalization}).
    \item We evaluate \name against state-of-the-art baselines and demonstrate its effectiveness (Section \ref{sec:evaluation}).
\end{itemize}

\section{Related Work}
\label{sec:related}

\noindent\textbf{NL2SQL.} 
Generating accurate SQL queries from natural language questions (NL2SQL) is a long-standing challenge due to the complexities in user question understanding, database schema comprehension, and SQL generation~\citep{DBLP:journals/ftdb/QuamarELO22,DBLP:journals/vldb/KatsogiannisMeimarakisK23,10.14778/3681954.3682003}. Recently, large language models (LLMs) have demonstrated significant capabilities in natural language understanding as the model scale increases. Many LLM-based NL2SQL solutions~\citep{DBLP:journals/corr/abs-2312-11242,10.1145/3589292,DBLP:journals/corr/abs-2405-16755,10.1145/3654930,10.14778/3681954.3682003} have emerged. 

SC-Prompt~\cite{10.1145/3589292} divides the NL2SQL task into two simpler sub-tasks (i.e., a structure stage and a content stage). Structure and content prompt construction and fine-grained constrained decoding are proposed to tackle these two sub-tasks, respectively.
CodeS~\cite{10.1145/3654930} adopts an incremental pre-training approach to enhance the SQL generation capability. It further addresses the challenges of schema linking and domain adaptation through prompt construction and data augmentation.
MAC-SQL~\cite{DBLP:journals/corr/abs-2312-11242} is an LLM-based multi-agent collaborative framework, in which a decomposer agent leverages few-shot chain-of-thought reasoning for SQL generation and two auxiliary agents utilize external tools or models to acquire smaller sub-databases and refine erroneous SQL queries.
CHESS~\cite{DBLP:journals/corr/abs-2405-16755} introduces a new pipeline that consists of effective relevant data/context retrieval, schema selection, and SQL synthesis. Also, CHESS is equipped with an adaptive schema pruning technique based on the complexity of the problem and the model's context size.

These methods focus on generating more accurate SQL queries for a given NL question. However, they neither explicitly handle schema ambiguity nor learn the preferences from users. Consequently, these state-of-the-art methods are not able to learn from user feedback and improve their SQL generation progressively.

\noindent\textbf{Ambiguity in SQL.}
While ambiguity has been extensively studied in many fields of NLP~\cite{li-etal-2022-visa,futeral-etal-2023-tackling}, it has not been explored much in NL2SQL. Hou et al.,~\cite{DBLP:conf/icml/HouLQAC024} introduce an uncertainty decomposition framework for LLMs in general, which can be applied to any pre-trained LLM. CAmbigNQ~\cite{lee-etal-2023-asking} focuses on ambiguous questions in open-domain question answering, tackling the problem in three steps, ambiguity detection, clarification question generation, and clarification-based QA. CLAM~\cite{DBLP:journals/corr/abs-2212-07769} is a framework that drives language models to selectively ask for clarification about ambiguous user questions and give a final answer after receiving clarification for open-domain QA. These works focus on clarifying the intent of the question, rather than resolving ambiguities within the data where the answer resides.

In NL2SQL, Wang et al.~\cite{wang-etal-2023-know} tackle ambiguity in SQL arising from related column names. The proposed method relies on labeling words of the text and does not generalize to other types of ambiguity beyond column ambiguity. 
AmbiQT~\cite{ambiqt} represents the first open benchmark for testing coverage of ambiguous SQLs. It introduces a decoding algorithm that searches the SQL logic space by combining plan-based template generation with a beam-search-based infilling. However, it does not capture and adapt to user preferences over the course of multiple questions.

\section{Problem Formulation}
\label{sec:overview:problem}

Traditional NL2SQL systems typically generate a single SQL query in response to a user's question. The main goal is for the generated SQL query to match the ground truth SQL query in terms of execution result. However, this approach may fall short in cases where the question contains ambiguity. In such instances, the generated query might not align with what the user intended. To address this, we propose developing an NL2SQL system that generates a set of possible queries, allowing the user to select the one that best meets their needs.

Given a user question \( Q \), the system generates a set of SQL queries to potentially answer the question. The process can be expressed as:
\begin{equation*}
\{ \text{SQL}_1, \dots, \text{SQL}_k \} = NL2SQL(Q)
\end{equation*}

We assume that although the natural language question input to the system contains semantic ambiguities, there is nonetheless a single ``correct'' SQL query which captures the user's true intent and preferences, which we refer to as the ground truth query \( SQL_{GT} \). The system aims to maximize the likelihood that the execution of any SQL query in the final set returns the same result as the execution of \( SQL_{GT} \). Let \( EX(SQL) \) denote the execution result of query \( SQL \). Let \( A(Q) \) denote the execution accuracy for question \( Q \), defined as:

\begin{equation*}
A(Q) = 
\begin{cases} 
1 & \text{if } \exists \, SQL \in \text{NL2SQL(Q) such that } EX(SQL) = EX(SQL_{GT}) \\
0 & \text{otherwise}
\end{cases}
\end{equation*}

The average accuracy over a workload of natural language questions \( W = \{Q_1, Q_2, \dots, Q_N\} \) is:

\begin{equation*}
\text{AvgAcc}(W) = \frac{1}{N} \sum_{i=1}^{N} A(Q_i)
\end{equation*}

In concept, generating all possible SQL queries for a question will achieve 100\% accuracy, but it is impractical as it overwhelms users with too many options. Maximizing accuracy alone does not guarantee a good user experience. When users are presented with an excessive number of choices, identifying the correct query becomes challenging. Therefore, it is essential to balance accuracy and the size of the result set to enhance user experience.

The optimization goal of the system is to maximize the likelihood of including the correct SQL query in a manageable set of results (\( K \)), ensuring efficiency and usability. The average number of SQL queries shown over the workload \( W \) is:

\begin{equation*}
\text{AvgResultSize}(W) = \frac{1}{N} \sum_{i=1}^{N} | NL2SQL(Q) |
\end{equation*}

The overall objective is to maximize average accuracy while limiting the average number of results.

\begin{equation*}
\textbf{max} \, \text{AvgAcc}(W) \text{,}\  \textbf{ subject to}\   \text{AvgResultSize}(W) \leq K
\end{equation*}

This balance is key for optimizing both system performance and user experience.

\section{\name System Overview}
\label{sec:odin_overview}
\begin{figure*}[t]
    \begin{center}
        \includegraphics[width=0.99\linewidth]{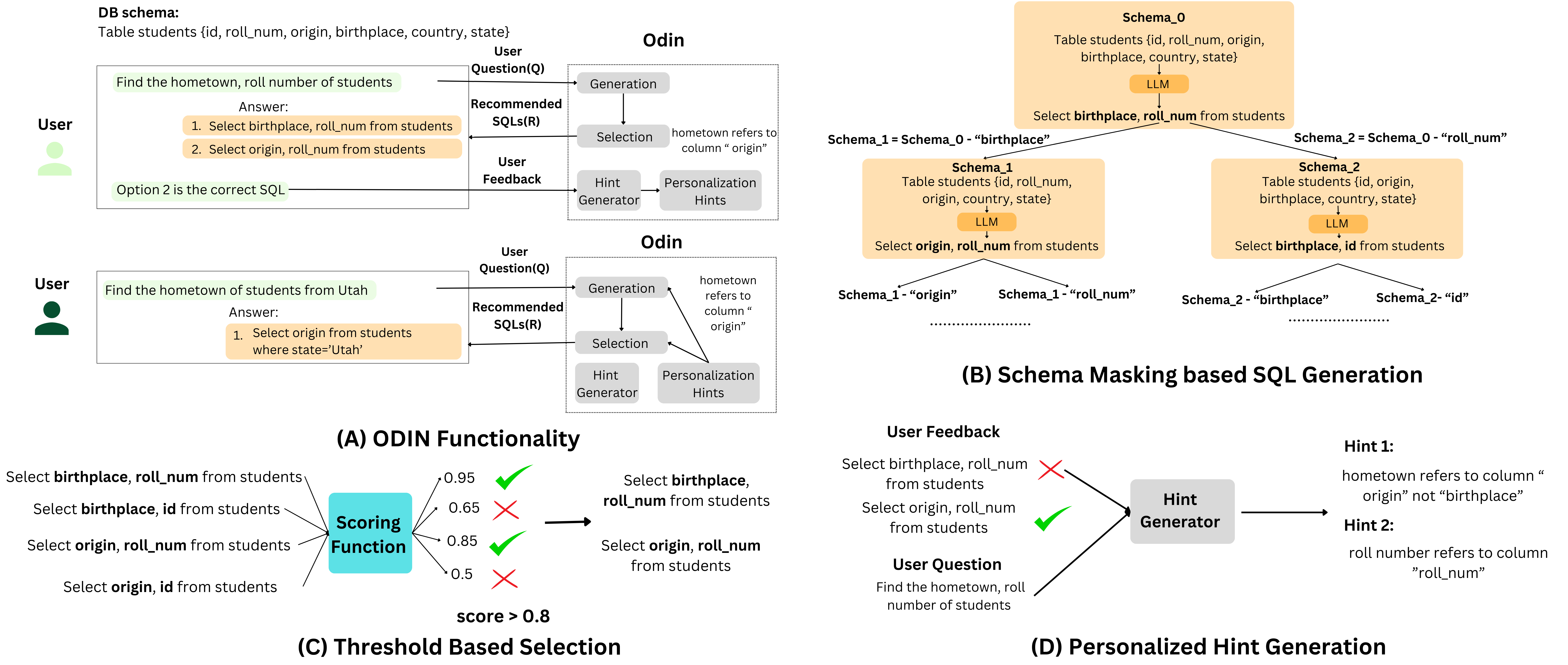}
    \end{center}
    \caption{\name Overview}
    \label{fig:odin_overview}
\end{figure*}

\name is an NL2SQL recommendation engine designed to assist users in managing schema ambiguities within their databases by generating multiple SQL query options and further resolving these ambiguities through learning user preferences. Upon receiving a natural language user question, \name generates several potential SQL queries, as demonstrated in \cref{fig:odin_overview}(A). \name personalizes query generation by incorporating user feedback. The core novelty of \name compared to other NL2SQL systems lies in its ability to generate a small, accurate set of SQL queries tailored to user preferences. Internally, \name is composed of three key components: the Generator, the Selector, and the Personalizer.

\noindent\textbf{Generator:} This component generates potential SQL candidates based on the user’s question. A typical approach used by other NL2SQL systems~\cite{wang2023mac,talaei2024chesscontextualharnessingefficient} for generating multiple SQL query candidates is to repeatedly call the LLM with the same prompt, using high temperatures and sampling techniques such as nucleus sampling~\cite{holtzman2019curious} to induce different output SQL queries with each call. However, \cite{ambiqt} shows that this approach does not produce sufficiently diverse SQL queries, because LLMs tend to produce simple variants of the same SQL query during sampling. For instance, the model might generate queries such as \texttt{select * from students}, \texttt{SELECT * FROM students}, or \texttt{select * from students;} with only minor cosmetic variations.

\namenospace's Generator improves on this baseline approach by incorporating previously-generated queries into the generation process. Specifically, our method selectively masks certain schema elements used in the previously-generated SQL queries, encouraging the LLM to explore alternative schema elements. For instance, as illustrated in Figure \ref{fig:odin_overview}(B), the initial SQL query is generated using the entire schema. Key columns, such as \texttt{birthplace} and \texttt{roll\_num}, are then marked as candidates for masking. By excluding these columns in subsequent LLM calls, new \textit{masked schemas} are generated, leading to different SQL queries. For example, in the newly generated SQL query, the \texttt{origin} column replaces \texttt{birthplace}, and \texttt{id} replaces \texttt{roll\_num}. Although this approach can produce diverse queries, it risks inefficiency due to the exponential number of possible masked schemas. This is particularly problematic when LLM calls are computationally expensive and should be minimized. Section \ref{sec:generation} describes our approach to generate diverse SQL queries while minimizing the number of LLM calls.

\noindent\textbf{Selector:} The generation algorithm may occasionally mask schema elements for which no reasonable substitutes exist, leading the LLM to produce incorrect SQL queries. Additionally, some generated queries might not align with user preferences. The primary objective of \name is to maintain a compact set of generated SQL queries while ensuring accuracy. Removing these incorrect and misaligned queries can help \name reduce the set size.

To address this issue, after \namenospace's Generator has produced a set of candidate SQL queries, \namenospace's Selector filters out candidate SQL queries which are likely to be incorrect, ensuring that nearly all correct SQL queries are preserved. To ensure accuracy in this filtering process, we employ the conformal prediction framework~\cite{conformaltutotrial}, which provides statistical guarantees for our ability to retaining correct queries. Specifically, we evaluate candidate SQL queries with a scoring function, selecting those exceeding a threshold (e.g., \cref{fig:odin_overview}(C) shows an example where candidates with scores above 0.8 are retained). The effectiveness of this process hinges on the scoring function and threshold selection, which we discuss further in \cref{sec:selection}.

\noindent\textbf{Personalizer:} Users may prefer that SQL queries use certain tables and columns over others, and learning these preferences can enhance recommendation quality. Since preferences are specific to the application, schema, or database, they require user input. These preferences must also be conveyed to the Generator and Selector components to personalize outputs.

After \name produces the set of SQL queries for a given user question, the user is asked to select the correct SQL query from the output set, if any. This feedback is transformed into textual hints, which the Generator and Selector use for personalization. For instance (\cref{fig:odin_overview}(D)), if a user selects a SQL query that uses the \texttt{origin} column instead of \texttt{birthplace}, the system infers that the term \textit{hometown} from the user’s query maps to \texttt{origin} rather than to \texttt{birthplace}. The hint generator then processes this feedback, along with the user’s current question, to perform schema linking (i.e., mapping specific entities from the user’s input question to corresponding schema elements). Through this schema linking, \name can map natural language phrases like \textit{hometown} and \textit{roll number} from the user’s question to the \texttt{origin} and \texttt{roll\_num} columns, respectively. These personalized hints are stored and later used by the Generator and Selector to align future SQL queries with user preferences. In \cref{fig:odin_overview}(A), when the user asks, ``\textit{hometown of students from Utah}'', the system uses the hint that \textit{hometown} refers to \texttt{origin}, enabling it to generate the correct SQL query. We describe the personalization process in further detail in \cref{sec:personalization}.

\section{Generator}
\label{sec:generation}
For a given question that can be answered with multiple possible schema components, \name ensures all potential SQL queries using these components are generated and presented to the user. The generator in \name handles this by producing the set of potential queries.  
A common method in LLM pipelines for generating diverse answers is using high-temperature sampling, where the temperature controls randomness. Higher temperatures lead to more varied and creative responses, while lower temperatures produce more focused, predictable results.

One might assume that high temperatures would generate diverse SQL queries, but in practice, it leads to only superficial differences, such as varying table/column aliases, using equivalent functions, or altering the SQL structure with subqueries. While these changes affect the query's appearance, they don't impact execution, resulting in only cosmetic diversity in generated SQL.

For example, consider the question: \emph{``List the names of customers who have placed orders over \$1,000 in the past six months.''} At a high temperature setting, the model might generate the following SQL in one run:

\begin{verbatim}
SELECT Name FROM Customers
JOIN Orders ON Customers.CustomerID = Orders.CustomerID
WHERE Orders.TotalAmount > 1000 
AND Orders.OrderDate >= DATE_SUB(CURDATE(), INTERVAL 6 MONTH);
\end{verbatim}

This SQL joins the \texttt{Customers} and \texttt{Orders} tables and filters for orders over \$1,000 within the last six months. On a second run, it might produce:

\begin{verbatim}
SELECT c.Name FROM Customers c
INNER JOIN Orders o ON c.ID = o.CustomerID
WHERE o.TotalAmount > 1000 
AND o.OrderDate >= DATE_SUB(CURDATE(), INTERVAL 6 MONTH);
\end{verbatim}

This version simply uses different aliases for the tables compared to the first query, but it will yield the same execution result. Finally, on a third run, the model might generate:

\begin{verbatim}
SELECT Name FROM Customers
WHERE CustomerID IN (
    SELECT CustomerID FROM Orders
    WHERE TotalAmount > 1000 
    AND OrderDate >= DATE_ADD(CURRENT_DATE, INTERVAL -6 MONTH)
);
\end{verbatim}

Although this query uses a subquery and a different date function, it yields the same output as the previous two queries. This example demonstrates how high temperatures in LLMs generate superficial variations in NL2SQL tasks without changing the effective SQL, as studied in \cite{ambiqt}.

Since high-temperature sampling produces only superficial diversity in results, we can instead prompt the LLM to generate SQL queries with different execution outcomes. To achieve this, we provide the LLM with all previously generated SQL queries and explicitly instruct it (via the LLM prompt) to create a new SQL query that yields a distinct execution result. We refer to this approach as \textit{ForcedDiversity}. Our empirical results show that \textit{ForcedDiversity} generates more truly diverse queries. However, it starts to repeat outcomes when too many SQL queries are included in the prompt.

To improve upon previous methods, \name leverages the structure of the NL2SQL task to enhance diversity by introducing controlled perturbation of schema information in the prompt. By masking certain parts of the schema used in earlier queries, the LLM is forced to rely on different schema elements, leading to more varied SQL queries.  For example, for the question \textit{``Find the hometown of students''}, if the database has both \texttt{birthplace} and \texttt{origin} columns, and the LLM initially uses \texttt{birthplace}, we can remove knowledge of this column's existence in subsequent prompts to encourage the LLM to generate a query using \texttt{origin}, leading to more diverse SQL queries.

We use the term \textit{masked schema} to refer to a subset of the full schema, i.e., a schema in which certain elements have been masked. We first present a simple algorithm for schema masking. We then use the limitations of this simple algorithm to motivate the schema masking algorithm used by \namenospace's Generator.

\subsection{Naive Schema Masking Algorithm}
\label{sec:generation:naive}
A naive version of the schema masking algorithm is shown in \cref{fig:odin_overview}(B). It generates SQL queries by progressively restricting a given schema and exploring different masked schemas in a tree-like structure. The root node of the tree structure represents the complete schema of the database. Each node in this tree represents a masked schema that is a subset of the complete schema. At each node, a SQL query is generated based on the current schema, which is then added to the set of SQL queries.
The algorithm then modifies the current schema of that tree node by removing individual columns used in the generated query, which produces new schemas to explore in child nodes. This process continues until all relevant masked schemas are exhausted. In \cref{fig:odin_overview}(B), the root node represents the complete schema, and an initial SQL query is generated from it. Columns such as \texttt{birthplace} and \texttt{roll\_num}, which are used in this query, are then removed, leading to modified schemas that are used to generate additional SQL queries. This approach ensures that the SQL queries generated by the child nodes are different from those of their parent nodes. This is straightforward to verify because the schemas used to generate child queries are missing at least one column present in the parent SQL query. However, the naive schema masking algorithm has two main drawbacks.

First, this algorithm might result in redundant explorations of the same masked schema. The algorithm does not ensure that descendants of a node explore distinct schemas. In some scenarios, the exploration paths of different branches may overlap. For example, in \cref{fig:odin_overview}(B), child schemas of the root node are generated by removing columns \texttt{birthplace} and \texttt{roll\_num}. The descendants of these child schemas might explore the same schemas in different orders (e.g., removing \texttt{birthplace} then \texttt{roll\_num}, vs. removing \texttt{roll\_num} then \texttt{birthplace}), leading to duplicated effort.

Second, the algorithm operates without accounting for practical resource constraints, such a constraint on the number of calls to an LLM. In real-world applications, the number of LLM calls is limited, but the exploration space of this algorithm grows exponentially with the size of the schema, leading to inefficient use of resources.

\subsection{\namenospace's Schema Masking Algorithm}
\label{sec:generation:final}

\namenospace's schema masking algorithm introduces two main improvements over the naive algorithm presented above: duplicate schema detection and a greedy exploration strategy for the search tree of schemas. The key idea behind duplicate schema detection is to maintain a record of schemas seen so far and to not explore redundant nodes.
To deal with the issue of limited resources, we introduce a greedy tree search strategy that uses a priority queue to explore nodes based on a scoring system. These scores reflect the relevance of schema of the current node to the entities mentioned in the user's question, allowing the algorithm to focus on the most promising exploration paths while staying within resource constraints.

\cref{alg:greedy_tree_search} shows the pseudocode for the algorithm. The algorithm takes the user question, database schema and a limit of LLM calls as input (Line 12). If the LLM fails to find relevant entities in the masked schema and is unable to generate a valid SQL query, the exploration may terminate early. The limit on LLM calls represents only the maximum number of attempts, not a guarantee that all calls will be used.
To prioritize exploration, the algorithm maintains a priority queue of nodes (i.e., schemas) to explore, where each node's priority is determined by its relevance score, computed by the \texttt{Cal\_Score} function (see \cref{sec:generation:scoring}). 
At each iteration, the node with the highest score is selected for exploration (Line 19).

When a node is explored, its corresponding schema is used to generate a SQL query, which is added to the results (Lines 20-22). Similar to the naive algorithm, new potential nodes (i.e., schemas) are generated by removing columns used in the SQL query (Lines 23-29). These new schemas are then scored using a relevance function. If the new schema has already been explored or is empty, it is discarded; otherwise, it is added to the priority queue. This ensures that the exploration is both resource-efficient and free from duplicate schemas.
In summary, the algorithm combines a greedy budget-conscious exploration strategy with a simple mechanism for avoiding redundant searches.

\subsection{Scoring Masked Schemas}
\label{sec:generation:scoring}

The schema scoring function, \texttt{Cal\_Score}, is crucial to the generation algorithm, as a noisy scoring mechanism can hinder efficient exploration. A masked schema's score intuitively reflects the likelihood that a plausibly-correct SQL query for the user's question can be formed from the elements of the masked schema. An \textit{entity} refers to information from the user's question, such as \textit{hometown} or \textit{roll number}. Each entity in the user's question should be well-represented by the masked schema in order to produce a SQL query. If a schema poorly represents any entity in the user's question, then the schema should intuitively get a lower score.

For example, consider the scenario in \cref{fig:odin_overview}(B). The entity \textit{hometown} could plausibly be represented by either the \texttt{birthplace} column or the \texttt{origin} column. In contrast, the entity \textit{roll number} can only be correctly represented by the column \texttt{roll\_num}. This distinction suggests that removing \texttt{birthplace} from the schema is less impactful than removing \texttt{roll\_num}. The \texttt{Cal\_Score} function reflects this intuition by computing the similarity between each entity in the question to its best-matching column in the schema and using the minimum similarity score across all entities as the schema's score.

In Algorithm \ref{alg:calc_score}, the function starts by extracting entities from the query (Line 10). It then calculates the maximum similarity score for each entity across all schema columns, storing these scores (Lines 11-18). The final schema score is determined by the lowest score among all entities, highlighting the schema’s weakest representation (Line 19). The \texttt{Extract\_Entities} function identifies entities in the user's question with an LLM call. The \texttt{Cal\_Sim} function, which measures entity-column similarity, utilizes SBERT similarity scores~\cite{reimers2019sentence}.

\begin{algorithm}
\begin{algorithmic}[1]
\caption{Greedy Tree Search with Resource Constraints}\label{alg:greedy_tree_search}
\State \textbf{Input:}
\State \hspace{1em} $f\_sch$ - Full schema
\State \hspace{1em} $q$ - Initial query
\State \hspace{1em} $max\_calls$ - Maximum number of LLM calls
\State \textbf{Output:}
\State \hspace{1em} $final\_queries$ - List of SQL queries

\State \textbf{Algorithm:}
\State \hspace{1em} $NL2SQL$ - Generates SQL for a given schema and user question
\State \hspace{1em} $Col\_Used$ - Returns columns used in a SQL query
\State \hspace{1em} $Remove\_Col$ - Removes a column from the schema
\State \hspace{1em} $Cal\_Score$ - Calculates relevance score for a schema
\Function{GenSQLQueries}{$f\_sch$, $q$, $max\_calls$}
    \State $final\_queries \gets []$
    \State $queue \gets \text{PriorityQueue}()$
    \State $schema\_seen \gets []$
    \State $queue.\text{push}((f\_sch, 1.0))$ \Comment{Initial score is 1}
    \State $llm\_calls \gets 0$
    \While{$queue$ \textbf{is not empty} \textbf{and} $llm\_calls < max\_calls$}
        \State $(curr\_sch, \_) \gets queue.\text{pop}()$
        \State $sql \gets NL2SQL(curr\_sch,q)$
        \State $llm\_calls \gets llm\_calls + 1$
        \State $final\_queries.\text{append}(sql)$
        \For{each $col$ \textbf{in} Col\_Used(sql)}
            \State $new\_sch \gets Remove\_Col(curr\_sch, col)$
            \State $score \gets Cal\_Score(new\_sch, query)$
            
            \If{$new\_sch$ \textbf{is not empty}}
                \If{$new\_sch$ \textbf{not in} $schema\_seen$}
                    \State $queue.\text{push}((new\_sch, score))$
                    \State $schema\_seen.add(new\_sch)$
                \EndIf
            \EndIf
        \EndFor
    \EndWhile
    
    \State \Return $final\_queries$
\EndFunction

\end{algorithmic}
\end{algorithm}

\begin{algorithm}
\begin{algorithmic}[1]
\caption{Calculate Schema Relevance Score}\label{alg:calc_score}
\State \textbf{Input:}
\State \hspace{1em} $schema$ - Current schema
\State \hspace{1em} $query$ - User question
\State \textbf{Output:}
\State \hspace{1em} $score$ - Relevance score of the schema
\State \textbf{Helper Functions:}
\State \hspace{1em} $Extract\_Entities$ - extracts entities from the user question
\State \hspace{1em} $Cal\_Sim$ - gives similarity score between an entity and a column

\Function{Cal\_Score}{$schema$, $question$}
    \State $entities \gets \text{Extract\_Entities}(question)$
    \State $entity\_scores \gets []$
    
    \For{each $entity$ \textbf{in} $entities$}
        \State $max\_similarity \gets -\infty$ 
        
        \For{each $col$ \textbf{in} $schema$}
            \State $similarity \gets \text{Cal\_Sim}(entity, col)$
            \If{$similarity > max\_similarity$}
                \State $max\_similarity \gets similarity$
            \EndIf
        \EndFor
        
        \State $entity\_scores.\text{append}(max\_similarity)$
    \EndFor
    
    \State $score \gets \text{min}(entity\_scores)$ 
    \State \Return $score$
\EndFunction

\end{algorithmic}
\end{algorithm}

Note that when computing the similarity between an entity and a column, considering the table leads to better semantic results. For example, replacing the \texttt{address} column from the \texttt{student} table with the \texttt{address} column from the \texttt{student\_registration} table may be more relevant than using the \texttt{address} column from the \texttt{teachers} table. Thus, we include table semantics when computing similarity as well.

\section{Selector}
\label{sec:selection}

The Generator may be overeager when masking schema elements, leaving no suitable schema elements for a given question entity, which can lead the LLM to select incorrect tables or columns or omit necessary elements, resulting in an inaccurate SQL query. Additionally, if user preferences for certain tables or columns are masked, the generated query may not align with their intent. As a result, queries produced by the Generator may be incorrect or misaligned. The Selector's primary role is to eliminate flawed SQL queries while ensuring the correct one is retained. We can express this objective as follows:

\begin{equation}
\begin{aligned}
\textbf{minimize} \quad & | \text{Sel}(\text{Gen}(Q)) |  \\
\textbf{subject to} \quad & \Pr \left( \text{EXM}(\text{Sel}\left(\text{Gen}(Q)\right), \text{SQL}^{gt}) = 1 \mid \right. \\
& \quad \left. \text{EXM}\left(\text{Gen}(Q)\right), \text{SQL}^{gt} = 1 \right) > (1 - \alpha)
\end{aligned}
\label{eq:proof}
\end{equation}

In this formulation, \( \text{EXM}(S, \text{SQL}^{gt}) \) checks if any query in \( S \) produces the same execution result as the ground truth query \( \text{SQL}^{gt} \). \( \text{Gen}(Q) \) is the set of SQL queries produced by the Generator for a query \( Q \), and \( \text{Sel}(\text{Gen}(Q)) \) is the filtered subset of queries produced by the Selector. The parameter \( \alpha \), typically between 1\% and 5\%, represents the acceptable margin of error. This ensures that the selected set retains at least one query matching the ground truth, with the probability exceeding \( 1 - \alpha \). The constraint minimizes the risk of discarding the correct query while reducing the number of SQL queries return to the user, balancing accuracy and efficiency.

The objective of the Selector closely aligns with the conformal prediction framework~\cite{angelopoulos2023conformal,conformaltutotrial}, which is commonly used in classification tasks. In conformal prediction, the goal is to select a subset of labels such that the true label is included with high probability. The key idea behind conformal prediction is straightforward: a scoring function is used to assess how likely a label is incorrect, and all labels below a carefully chosen threshold are selected. This threshold is determined using a calibration set, ensuring that the high-probability guarantees hold, assuming new queries come from the same distribution as the calibration set. Similarly, we aim to select a subset of SQL queries such that the correct one is retained with high probability. We provide an overview of conformal prediction in \cref{sec:selection:conformal}, explain how we map our problem to this framework in \cref{sec:selection:mapping}, and describe our scoring function in \cref{sec:selection:scoring_functions}.

\subsection{Background on Conformal Prediction}
\label{sec:selection:conformal}
Originally proposed by~\cite{vovk2005line}, conformal prediction provides a way to turn heuristic notions of uncertainty from machine learning models into rigorous sets that are guaranteed to contain the true outcome with a specified probability. This framework can be applied to both regression and classification tasks, making it highly adaptable for various machine learning applications. The central idea of conformal prediction is to construct a set of possible outcomes for a new input \( X_{\text{test}} \) such that the set will contain the true outcome \( Y_{\text{test}} \) with a user-specified confidence level \( 1 - \alpha \). The procedure for generating these prediction sets is non-parametric and relies on a calibration dataset to quantify the uncertainty of the model's predictions.

The conformal prediction framework proceeds through the following key steps:

\noindent\textbf{1. Heuristic Uncertainty Estimate:} Start with a pre-trained model that provides a score function \( s(x, y) \in \mathbb{R} \) for each input-output pair \( (x, y) \). This score function reflects the level of disagreement between the input \( x \) and the predicted or true output \( y \). A higher score indicates worse agreement between \( x \) and \( y \), i.e., more uncertainty.
    
\noindent\textbf{2. Calibration:} To quantify uncertainty, the model is first calibrated using a set of observed data points \( \{(X_1, Y_1), \dots, (X_n, Y_n)\} \). For each pair \( (X_i, Y_i) \), the score \( s(X_i, Y_i) \) is computed, yielding a set of calibration scores \( s_1 = s(X_1, Y_1), \dots, s_n = s(X_n, Y_n) \). A quantile threshold \( \hat{s} \) is then selected such that
    \begin{equation}
    \hat{s} = \left(\frac{\lceil (1 + n)(1 - \alpha) \rceil}{n}\right) \text{-th quantile of } \{s_1, \dots, s_n\}.
    \label{eq:quantile}
\end{equation}
This quantile ensures that future prediction sets will contain the true outcome with probability at least \( 1 - \alpha \).

\noindent\textbf{3. Prediction Set Formation:} For a new input \( X_{\text{test}} \), the score function is evaluated for various possible outputs \( y \). The prediction set \( C(X_{\text{test}}) \) is then formed by including all outputs \( y \) such that the score function satisfies:
\[
C(X_{\text{test}}) = \{ y : s(X_{\text{test}}, y) \leq \hat{s} \}.
\]
This ensures that the prediction set contains all outcomes where the model's uncertainty (as captured by the score function) is below the threshold \( \hat{s} \).

\begin{theorem}[Conformal Prediction Guarantee, \cite{vovk2005line}]
Given a calibration dataset \( \{(X_1, Y_1), \dots, (X_n, Y_n)\} \) drawn i.i.d. from the same distribution as the test point \( (X_{\text{test}}, Y_{\text{test}}) \), conformal prediction constructs a prediction set \( C(X_{\text{test}}) \) using the threshold in \cref{eq:quantile} for the true outcome \( Y_{\text{test}} \). The prediction set satisfies the following coverage guarantee:
\[
P \left( Y_{\text{test}} \in C(X_{\text{test}}) \right) \geq 1 - \alpha,
\]
where \( \alpha \) is a user-specified significance level. This guarantee holds regardless of the underlying distribution of the data.
\end{theorem}

\subsection{Mapping Selector to Conformal Prediction }
\label{sec:selection:mapping}

In this subsection, we demonstrate how the Selector aligns with the conformal prediction framework. By establishing this connection, we leverage the statistical guarantees provided by conformal prediction to ensure that the correct SQL query is retained with high probability after the selection process. Conformal prediction constructs prediction sets that contain the true outcome with a specified confidence level \(1 - \alpha\). To map our problem to this framework, we define the following components:

\noindent\textbf{Inputs and Outputs:} In our setting, the input \( X \) is the user's natural language query \( Q \), and the output \( Y \) is a candidate SQL query \( q \) generated by the language model (LLM).

\noindent\textbf{Score Function:} We define a score function \( s(Q, q) \) that quantifies the likelihood of the SQL query \( q \) being \emph{incorrect} for the given input \( Q \). A higher score indicates a higher chance that \( q \) is incorrect. The specific design of this score function is detailed in \cref{sec:selection:scoring_functions}.

\noindent\textbf{Calibration Set:} The calibration set consists of pairs \( \{ (Q_i, \text{Gen}(Q_i)) \} \), where \( Q_i \) are natural language queries and \( \text{Gen}(Q_i) \) are the corresponding sets of SQL queries generated by \name. Importantly, we only look at pairs for which a query with the same execution result to that of the correct SQL query \( \text{SQL}^{gt}_i \) is present in \( \text{Gen}(Q_i) \). For each pair, we compute the score \( s(Q_i, q) \) for all queries \( q \in \text{Gen}(Q_i) \), focusing on the score distribution for correct SQL queries.

\noindent\textbf{Quantile Threshold:} We determine the threshold \( \hat{s} \) by analyzing the scores of correct SQL queries in the calibration set, specifically using the quantile defined in \cref{eq:quantile}.

\noindent\textbf{Prediction Set Formation:} For a new input \( Q_{\text{test}} \), we generate a set of candidate SQL queries \( \text{Gen}(Q_{\text{test}}) \). We compute the scores \( s(Q_{\text{test}}, q) \) for each \( q \in \text{Gen}(Q_{\text{test}}) \). The selector then forms the prediction set \( \text{Sel}(\text{Gen}(Q_{\text{test}})) \) by including all queries with scores less than or equal to \( \hat{s} \):
\[
\text{Sel}(\text{Gen}(Q_{\text{test}})) = \{ q \in \text{Gen}(Q_{\text{test}}) \mid s(Q_{\text{test}}, q) \leq \hat{s} \}.
\]
Mapping the selector to the conformal prediction framework provides the coverage guarantee from \cref{eq:proof}. The effectiveness depends on the quality of score function \(s(Q, q)\) (see \cref{sec:selection:scoring_functions} for details).

\subsection{Scoring Functions} 
\label{sec:selection:scoring_functions}
In the Selector, the scoring function evaluates each SQL query generated by the model. The goal is to assign \emph{low scores} to correct SQL queries and \emph{high scores} to incorrect ones. Ideally, the ground truth SQL query would get a score of zero, while all others get higher scores, allowing for confident elimination of incorrect queries. However, designing a perfect scoring function is difficult due to the inherent complexity of natural language and SQL semantics.

The scoring function should capture the \emph{semantic alignment} between the user's question and the SQL query, focusing on the relevant tables and columns. To achieve this, we leverage the capabilities of language models to understand and represent semantic relationships. We develop two scoring functions: an \emph{LLM-based function} that directly evaluates the SQL query using an LLM, and an \emph{SBERT-based function} that heuristically decomposes the task and uses Sentence-BERT (SBERT)~\cite{reimers2019sentence} to compute semantic similarities. Each scoring function is described below.

\subsubsection{LLM-based Scoring Function}
To develop the LLM-based scoring function, we utilize the language model's ability to understand complex instructions by prompting it to evaluate whether a given SQL query correctly answers the user's question. One method is to ask the LLM to assign a score or probability directly \cite{xiong2023can}, while another is to use the logit probabilities of specific tokens. We adopt the logit probabilities approach for finer-grained scoring.

\begin{figure}[h] \centering \includegraphics[width=0.9\linewidth]{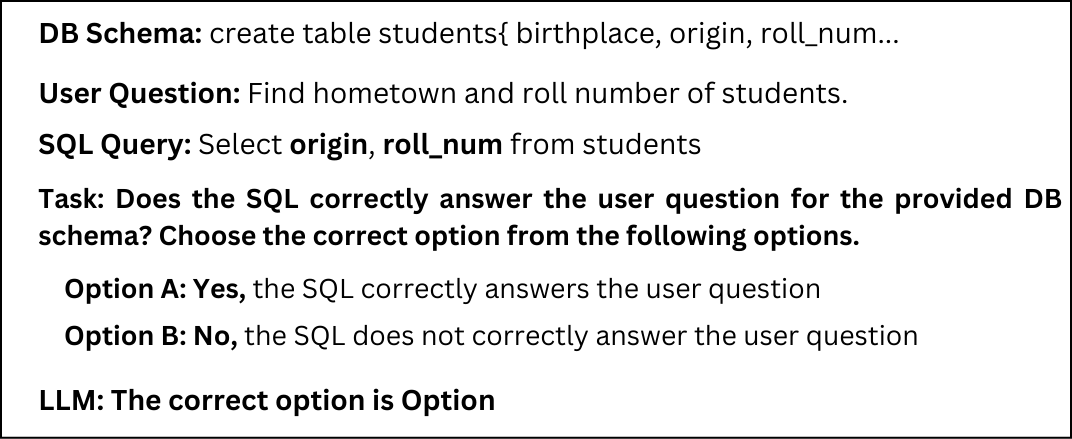} \caption{Prompt template used to score SQL queries using LLMs} \label{fig:llm_socring_template} \end{figure}

We use a prompt that asks the LLM if the SQL query answers the question, providing two options: \emph{A.} Yes, and \emph{B.} No (as shown in Fig.\ref{fig:llm_socring_template}). We use the logit probability of option \emph{B} as the score. A lower probability for \emph{B} indicates a higher likelihood that the SQL query is correct, aligning with our goal of assigning lower scores to correct queries.

\subsubsection{SBERT-based Scoring Function}
Although LLMs perform well, they can be resource-intensive. As a more efficient alternative, we propose an SBERT-based scoring function. SBERT computes semantic similarities between sentences or phrases \cite{reimers2019sentence}, which we leverage by breaking down the SQL query scoring task into smaller sub-tasks involving entity similarity computations.

Our approach is based on the observation that a correct SQL query should represent all entities mentioned in the user's question via its tables and columns, similar to the schema scoring function used in generation. The SBERT-based scoring follows \cref{alg:calc_score}, but focuses on the columns in the SQL query rather than the masked schema. By negating this score, we assign lower scores to queries that better represent the user's question. Queries missing entities receive higher scores, allowing the selector to filter them out. This method efficiently captures semantic alignment and is suitable for resource-limited scenarios.

\section{Personalization}
\label{sec:personalization}
As mentioned earlier, ambiguity in NL2SQL often arises when a database contains multiple similarly named tables or columns. For example, if a user asks, \textit{What were the total sales last year?}, but the database includes both \texttt{gross\_sales} and \texttt{net\_sales}, it is unclear which column the user intends to reference. User preferences are reflected in their choice of specific schema components. One user might consistently prefer the \texttt{gross\_sales} column when referring to total sales, while another may focus on \texttt{net\_sales}. These preferences are often tied to how the user phrases their query. For instance, one user might use the term \textit{total sales} to mean \texttt{gross\_sales}, while another might mean \texttt{net\_sales} when referring to \textit{final sales}. It is crucial to map the user's phrases or entities to the corresponding schema components.

Capturing user preferences can improve future recommendation, like in the previous example, if we know that a user associates \texttt{gross\_sales} with \textit{total sales}, future queries can avoid incorrectly selecting the \texttt{net\_sales} column in similar situations. While user preferences improve recommendation quality, the challenge lies in capturing them effectively. Without explicit feedback, it is difficult for the system to infer such preferences. Moreover, even after preferences are captured, integrating them into future query generation is not straightforward, as it requires biasing the system towards them. The Personalizer component addresses these challenges.

First, users select the correct SQL query from a set of displayed options, and this feedback forms the foundation for personalization in \namenospace. This feedback is then transformed into a format that influences both the Generator and Selector components of the system. Specifically, we provide these preferences in a textual format that biases the LLM's output. The key information conveyed in the textual hints is that when a user refers to a particular entity, a specific schema component should be chosen over alternatives. \name generates these hints by mapping the entities mentioned in the user's question to the corresponding schema components in the selected SQL query. Through these techniques, \name learns to associate entities in user questions with selected schema components, ensuring future queries are personalized based on past feedback.

\subsection{Generating Textual Hints}
\label{sec:personalization:hint}
Textual hints are used to guide both the Generator and Selector towards user preferences. The key information that textual hints capture is that when a user references an entity in their question, they prefer a specific schema components. We frame this as learning the correct schema linking based on the user's question and their preferred SQL query. Given a question $Q$, the correct SQL query ($SQL_{\text{True}}$), and a set of incorrect queries ($SQL_1, \dots, SQL_T$), the task is to map entities $E_1, \dots, E_P$ in the question to schema components in $SQL_{\text{True}}$ and the incorrect queries. Incorrect mappings help reduce the importance of wrong components, guiding the system towards accurate ones.

\begin{sloppypar}
For each entity \( E_i \), \name learns its mapping to the schema component in \( SQL_{\text{True}} \). For instance, if \( E_1 \) refers to \textit{total sales}, and \( SQL_{\text{True}} \) maps it to the \texttt{gross\_sales} column from the \texttt{customer\_sales} table, while incorrect queries map it to \texttt{net\_sales}, \name generates the following textual hint: \textit{When referring to \textbf{total sales}, the user prefers the \textbf{customer\_sales.gross\_sales} column over \textbf{customer\_sales.net\_sales}.}
\end{sloppypar}

The key subroutine for generating textual hints is schema linking, which maps entities from the user's question to the relevant tables and columns. Schema linking is a well-studied problem with approaches using general LLMs~\cite{scprompt}, specialized LLMs~\cite{li2024codes}, and SBERT models~\cite{wang2021ratsql,fu2023catsql}. While any method could be used, we propose a heuristic SBERT-based algorithm for greater efficiency.

The algorithm for generating textual hints, as detailed in \cref{alg:schema_linking}, links entities from the user's question to schema components and creates textual hints summarizing the correct mappings. This process begins with three inputs: the user’s question, the correct SQL query, and a set of incorrect SQL queries. The output is a list of textual hints based on the correct schema mappings.

The core function, \texttt{GenerateHints}, starts by extracting entities from the user’s question using the \texttt{Extract\_Entities} function. For each entity, the \texttt{SchemaMap} function determines the most likely schema component in the correct SQL query. Simultaneously, the algorithm gathers mappings from incorrect SQL queries and compares them with the correct mappings, recording any discrepancies as incorrect mappings. The \texttt{SchemaMap} function computes the similarity between an entity and each column in the SQL query using \texttt{Cal\_Sim}, selecting the column with the highest similarity score as the correct mapping. This ensures that each entity is accurately linked to the most relevant schema component, enhancing the precision of the generated hints.

Finally, the \texttt{Format\_Hint} function integrates the entity, the correct schema mapping, and the list of incorrect mappings into a template for generating hints. After processing all entities, the function returns the list of textual hints. 

\begin{algorithm}
\begin{algorithmic}[1]
\caption{Schema Linking and Hint Generation}\label{alg:schema_linking}
\State \textbf{Input:}
\State \hspace{1em} $Q$ - User question
\State \hspace{1em} $SQL_{\text{True}}$ - Correct SQL query
\State \hspace{1em} $SQLs_{\text{Incorrect}}$ - List of incorrect SQL queries
\State \textbf{Output:} 
\State \hspace{1em} $hints$ - List of textual hints for each entity
\State \textbf{Helper Functions:}
\State \hspace{1em} $Extract\_Entities$ - extracts entities from the user question
\State \hspace{1em} $Cal\_Sim$ - gives similarity score between an entity and a column
\State \hspace{1em} $Format\_Hint$ - generates textual hint using a template and provided entity ans schema components. 

\Function{GenerateHints}{$Q$, $SQL_{\text{True}}$, $SQLs_{\text{Incorrect}}$}
    \State $entities \gets \text{Extract\_Entities}(Q)$
    \State $hints \gets []$

    \For{each $entity$ \textbf{in} $entities$}
        \State $correct\_map \gets \text{SchemaMap}(entity, SQL_{\text{True}})$
        \State $incorrect\_map \gets []$
        
        \For{each $SQL$ \textbf{in} $SQLs_{\text{Incorrect}}$}
            \State $mapping \gets \text{SchemaMap}(entity, SQL)$
            \If{$mapping \neq correct\_map$}
                \State $incorrect\_map.\text{append}(mapping)$
            \EndIf
        \EndFor

        \State $hint \gets \text{FormatHint}(entity, correct\_map, incorrect\_map)$
        \State $hints.\text{append}(hint)$
    \EndFor
    
    \State \Return $hints$
\EndFunction

\Function{SchemaMap}{$entity$, $SQL$}
    \State $max\_sim \gets -\infty$
    \State $best\_mapping \gets None$

    \For{each $col$ \textbf{in} $SQL$}
        \State $sim \gets \text{Cal\_Sim}(entity, col)$
        \If{$sim > max\_sim$}
            \State $max\_sim \gets sim$
            \State $best\_mapping \gets col$
        \EndIf
    \EndFor

    \State \Return $best\_mapping$
\EndFunction
\end{algorithmic}
\end{algorithm}

\subsection{Using Personalization Hints}
\label{sec:personalization:using_hint}
\begin{figure}[h] 
\centering \includegraphics[width=0.9\linewidth]{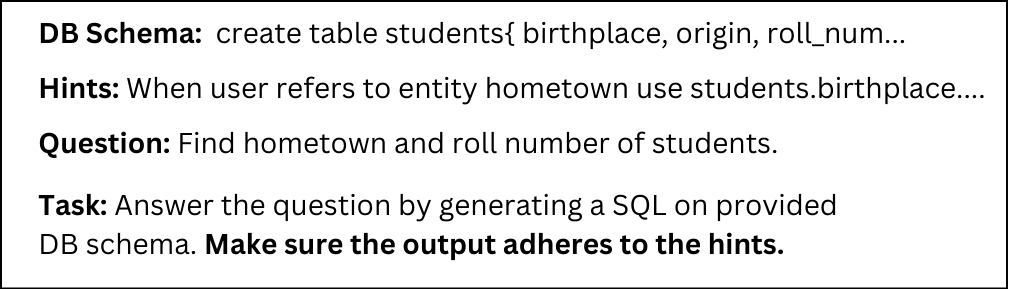} \caption{Prompt template used to provide hints to LLM} \label{fig:hint_template}
\end{figure}

We leverage LLMs' in-context learning to incorporate personalization hints during both the Generator and Selector stages. In the Generator stage, hints guide SQL query generation by including them in the LLM's context, as shown in Figure~\ref{fig:hint_template}. For the Selector, if using an LLM-based scoring function, hints are similarly added to the LLM’s context. However, with an SBERT-based scoring function, integrating textual hints is less straightforward. To address this, we fine-tune the SBERT model by adjusting its representations, ensuring that preferred schema components align closely with the corresponding entities, while non-preferred components are pushed further away. This is done using triplets of entities and schema components, such as fine-tuning SBERT to ensure \textit{total sales} is closer to \texttt{gross\_sales} and farther from \texttt{net\_sales}.

\subsection{Discussion}
\begin{sloppypar}
User preferences can evolve over time, known as preference drift. For example, a user might initially associate \textit{total sales} with \texttt{gross\_sales}, but later shift to \texttt{final\_sales}. Such preference drift can reduce the quality over time. Detecting and adjusting for preference drift is a well-studied in recommender systems literature. A common approach to address this is by maintaining a sliding window of recent user feedback~\cite{concept_drift}, allowing the system to adapt to the most recent interactions. Alternatively, we can use decay functions to gradually reduce the influence of older preferences~\cite{decay_preference}. Also, human-in-the-loop systems can be employed to manually adjust preferences~\cite{hil_preference_drift}.
\end{sloppypar}

User preferences can sometimes be weak, where a user typically favors alternative X (e.g., \texttt{gross\_sales}) but occasionally prefers Y (e.g., \texttt{net\_sales}). In such cases, it is best to present both alternatives. A technique that enforces strong preferences would only display X, being the favored option. However, \name can adapt to both scenarios. In case of a strong preference, \name can learn to exclude Y, as only X would be marked correct in the calibration set. If it is weakly enforced, both X and Y will be correct in the calibration set. While X may receive a higher score during the selector stage, the cut-off will be set such that both X and Y are included in the final selection, as either could be correct.

\section{Evaluation}
\label{sec:evaluation}
We begin by detailing the experimental setup (\cref{sec:evaluation:setup}) and then present the results of an extensive study comparing \name with various baseline methods. The evaluation reveals several key findings:
\begin{itemize}
    \item \name consistently produces a higher-quality SQL result set for user recommendations compared to the baselines. The result sets are not only smaller but also include the correct SQL query more frequently (\cref{sec:evaluation:overall}).
    \item Even without personalization, \name outperforms the baselines due to its Generator and Selector components (\cref{sec:evaluation:overall}).
    \item The schema masking-based Generator in \name generates a more diverse range of SQL queries compared to both generic and NL2SQL-specific diversity-inducing techniques (\cref{sec:evaluation:generator}).
    \item \namenospace's Selector significantly reduces the size of the result set while retaining the correct SQL query. The LLM-based scoring used in this stage proves more effective than SBERT-based scoring (\cref{sec:evaluation:selector}).
\end{itemize}

\subsection{Experimental Setup}
\label{sec:evaluation:setup}

\begin{figure}[h] 
\centering \includegraphics[width=\linewidth]{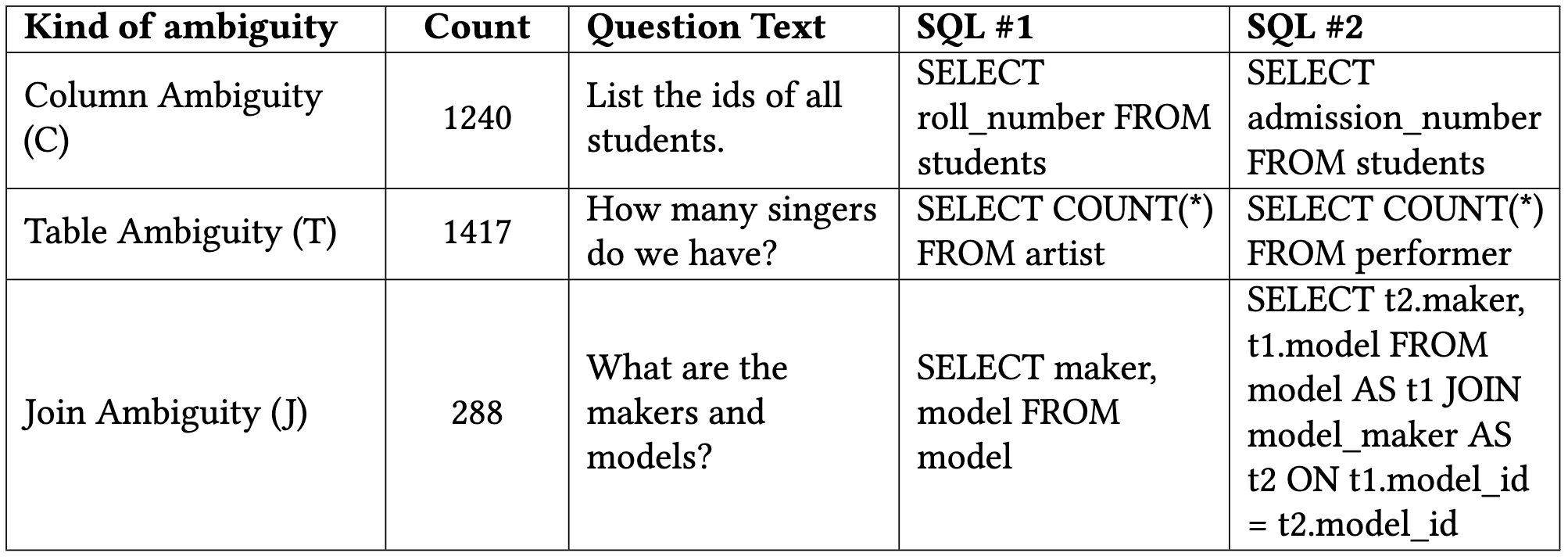} \caption{Different types of ambiguities in the AmbiQT Benchmark.} \label{fig:ambiqt_desc}
\end{figure}

\noindent \textbf{Datasets}. We use two benchmarks to evaluate our \name system.
\begin{itemize}
    \item \textbf{AmbiQT Benchmark}: This benchmark is a modified version of the Spider benchmark \cite{yu2019spider} and addresses different types of ambiguities commonly found in databases (see \cref{fig:ambiqt_desc}). In each case, the benchmark modifies the database schema such that there are two correct SQL statements for each question. We use this benchmark to evaluate the Generator component in \namenospace.
    \item \textbf{Mod-AmbiQT Benchmark}: The original AmbiQT benchmark is not ideal for evaluating personalization, so we created a modified version, Mod-AmbiQT. In this new benchmark, we introduce duplicate columns and tables based on the AmbiQT benchmark.

    Entities in the questions can map to either of the alternatives, but only one mapping is considered correct. As a result, each question in this database can now have between 2 to 8 plausibly-correct SQL queries, depending on the number of entities in the question, although only one specific SQL query is designated as correct. This modification is useful for testing personalization.

    For example, in the \texttt{concert} database, we introduce \texttt{artist} and \texttt{performers} as two alternatives for the singer table. For the question ``\textit{How many singers do we have?}'', the two alternative SQL statements could be: ``\texttt{SELECT COUNT(*) FROM artists}'' and ``
    \texttt{SELECT COUNT(*) FROM performers}''. Out of these, only the first query is considered correct. The entity \textit{singers} consistently maps to \texttt{artist} across all questions in the database, allowing us to evaluate personalization.
    
    The new benchmark includes 1,298, 2,148, and 626 Q/A pairs for table, column, and join ambiguity, respectively. We use this benchmark for overall system evaluation.
\end{itemize}

\noindent \textbf{Baselines.} We compare \name against two main baselines.
\begin{itemize}
    \item \textbf{Diverse Sampling}: In this baseline, we set the LLM temperature to a high value (\textit{temp}=1.0) to promote the generation of diverse SQL queries. The resulting SQL queries are then presented to the user.
    
    \item \textbf{Forced Diversity}: In this baseline, we provide the LLM with all previously generated SQL queries and instruct it to produce a new SQL query that differs from the prior ones. The generated SQL queries are then shown to the user.
    
    \item \textbf{\namenospace}: The \name baseline comprises three modules: \textit{Generator}, \textit{Selector}, and \textit{Personalizer} enabled by default, with various components enabled or disabled based on the specific variant. In the Selector, a LLM-based scoring function is employed unless otherwise specified.
\end{itemize}

All the above baselines are model-agnostic, meaning any LLM can be used. For our evaluation, we utilize Claude 3 Haiku.

To further evaluate the effectiveness of \namenospace's Generator, we compare it against six baselines from \cite{ambiqt} that aim to induce diversity in NL2SQL generation. The goal is to capture all possible ambiguous SQL queries for a given question.

First, we consider a set of naive baselines based on pre-trained language models (PLMs) and LLMs, showcasing their limited ability to generate diverse SQL statements. 

\noindent\textbf{{1. LLM-X}}{: a commercially available LLM specialized for coding tasks.}

\noindent\textbf{{2. RESDSQL~\cite{li2023resdsql}}}: one of the top-performing methods on the SPIDER benchmark. We use the 3B variant of RESDSQL, which is the most powerful, for comparison purposes. However, we disable the representation from \cite{gan2021natural}, as it is orthogonal to our approach and could be used alongside it.

Next, we examine baselines that incorporate common sampling techniques to promote more diverse generation. For this, we use the T5-3B checkpoint from the PICARD \cite{scholak2021picard} repository, which fine-tunes T5-3B on the SPIDER dataset. These baselines include:

\noindent\textbf{3. Beam Search (T5-3B-beam)}: We apply Beam Search with a width of 10 as the default decoding strategy for T5-3B.
    
\noindent\textbf{4. Top-k Sampling (T5-3B-k)}: At each decoding step, we sample from the top-50 tokens using top-k sampling with \(k = 50\).
    
\noindent\textbf{5. Nucleus/Top-p Sampling (T5-3B-p)}: We apply top-p sampling at each decoding step, where tokens that account for 90\% of the probability mass are considered, as proposed in \cite{holtzman2019curious}.

\noindent\textbf{6. Logical Beam}: we include this decoding algorithm that navigates the SQL logic space by combining plan based template generation and constrained infilling. For this, we fine-tune a version of Flan T5-3B (with a maximum input length of 512) using the Adafactor optimizer \cite{shazeer2018adafactor} (learning rate $10^{-4}$, no decay) on the AmbiQT benchmark, as described in \cite{ambiqt}.

$\ \ $\\
\noindent \textbf{Metrics:}
To evaluate the system, we use two key metrics: average accuracy over the workload ($AvgAcc$) and average number of results shown to the user ($AvgResultSize$), both defined in \cref{sec:overview:problem}. $AvgAcc$ checks if there exists a SQL query within the result set that matches the execution result of the golden SQL query, and computes the average accuracy over the entire workload. On the other hand, $AvgResultSize$ measures the average number of SQL queries presented to the user.

\subsection{Overall Evaluation}
\label{sec:evaluation:overall}

\begin{figure*}[t]
    \begin{center}
        \includegraphics[width=\linewidth]{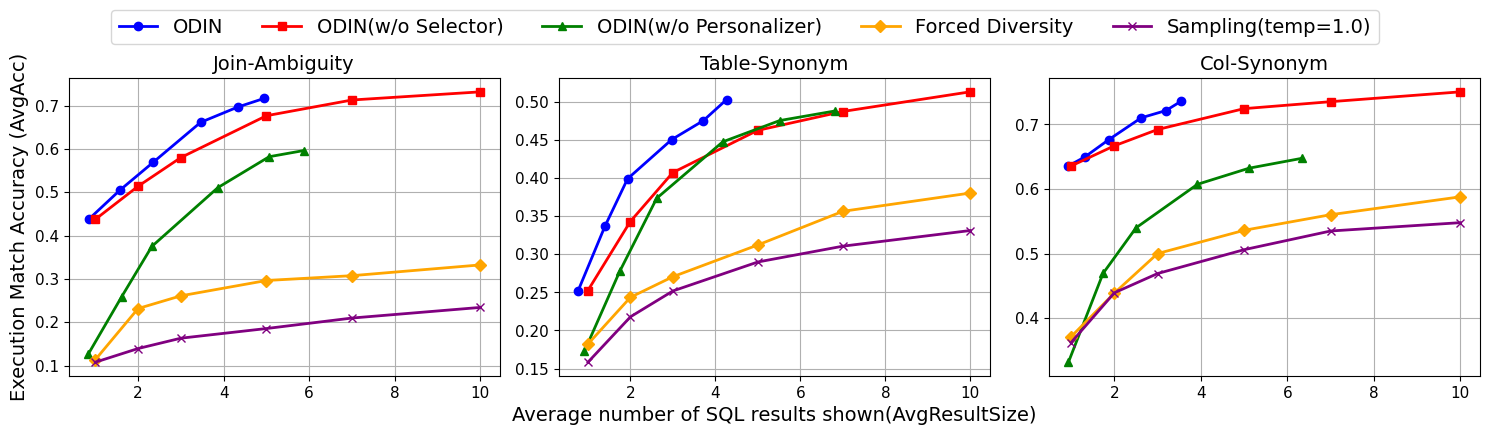}
    \end{center}
    \caption{Execution Match Accuracy ($AvgAcc$) versus the average number of results shown to the user ($AvgResultSize$) across three different ambiguity types for various baselines. \name can achieve up to twice the accuracy while presenting only half the number of SQL queries to the user compared to the next best baseline.
    }
    \label{fig:expt_1}
\end{figure*}

We first examine how \name improves accuracy across different types of ambiguities. For this experiment we use the Mod-AmbiQT benchmark. \cref{fig:expt_1} illustrates the average accuracy of the workload ($AvgAcc$) of the generated results versus the average number of SQL results shown to the user ($AvgResultSize$) for three distinct ambiguity types: join ambiguity, table ambiguity, and column ambiguity. The figure compares \name with two baselines: Sampling and Forced Diversity.

For all methods, we gradually increase the number of LLM calls $K$ (specifically, we set $K$ to 1, 2, 3, 5, 7, and 10) used for generating SQL queries. As the budget increases, each method shows a higher number of SQL queries to the user, which improves accuracy. Note that for \namenospace, $K$ refers to the number of LLM calls used in the Generator, i.e., the Generator produces $K$ SQL queries. However, $K$ does not include LLM calls made by the Selector and Personalizer.

\noindent\textbf{Overall Performance:} \name consistently outperforms all other methods, followed by its variants, then Forced Diversity, and finally Sampling. \name achieves accuracies of 71.6\%, 50.3\%, and 73.6\% for a budget of 10 LLM calls, while displaying an average of 4.9, 4.2, and 3.9 SQL queries to the user for join, table, and column ambiguities, respectively. In contrast, Forced Diversity achieves 33.2\%, 38.0\%, and 58.8\% accuracy across the workload. Compared to Forced Diversity, \name shows improvements of 38\%, 12\%, and 15\% in accuracy across the three ambiguity types while returning 2-3$\times$ fewer SQL queries. The improvement in accuracy is mainly due to the Generator and Personalizer components, while reduction in SQL result set size is attributed to the Selector component. Thus, \namenospace’s result set has up to twice the chance of containing the correct SQL result while being 2-3$\times$ smaller in size than Forced Diversity.

\noindent\textbf{Impact of Personalization:} We assess the impact of the Personalizer feature, which enables \name to learn user preferences and improve performance even at \( K = 1 \), as shown in Figure~\ref{fig:expt_1}. At \( K = 1 \), \name with Personalizer significantly outperforms \name without Personalizer by 30\%, 8\%, and 30\% for join, table, and column ambiguities, respectively. This improvement is due to personalized variants better understanding user preferences and generating correct solutions on the first attempt. Personalization can increase the likelihood of finding the correct SQL query in the result set by up to 4-5$\times$ for small result set sizes.

\noindent\textbf{Performance Without Personalization:} Despite the advantages of Personalizer, \name still performs effectively without it. For \( K = 1 \), the accuracy of \name without Personalization is comparable to that of the baselines. However, as \( K \) increases, \name without Personalization quickly surpasses these baselines. This behavior is expected, as with only one LLM call, all non-personalized baselines generate the same initial SQL query. With a budget of 10 LLM calls, \name without Personalization outperforms the baselines by 27\%, 10\%, and 8\%. Thus, although \name without Personalization starts with similar accuracy to the baselines for small result sets, it rapidly improves, achieving up to a 25\% gain in accuracy as the set size increases.

\noindent\textbf{Impact of the Selector:} We assess the Selector component, which is designed to enhance precision by filtering out incorrect SQL statements while keeping the correct ones, thereby improving precision without significantly affecting recall. Comparing \name with and without the Selector, we observe that both maintain similar accuracy, with the Selector reducing accuracy by only 1.6\%, 1\%, and 1.4\% for the respective ambiguity types. However, \name with the Selector displays an average of 5, 4.1, and 3.8 SQL queries per ambiguity type, compared to 10 queries generated by \name without the Selector. Thus, the Selector component significantly reduces the number of SQL queries displayed by up to 2-2.5$\times$ with only a marginal loss in accuracy.

\subsection{Effectiveness of Generator}
\label{sec:evaluation:generator}

\begin{figure*}[h]
    \begin{center}
        \includegraphics[width=\linewidth]{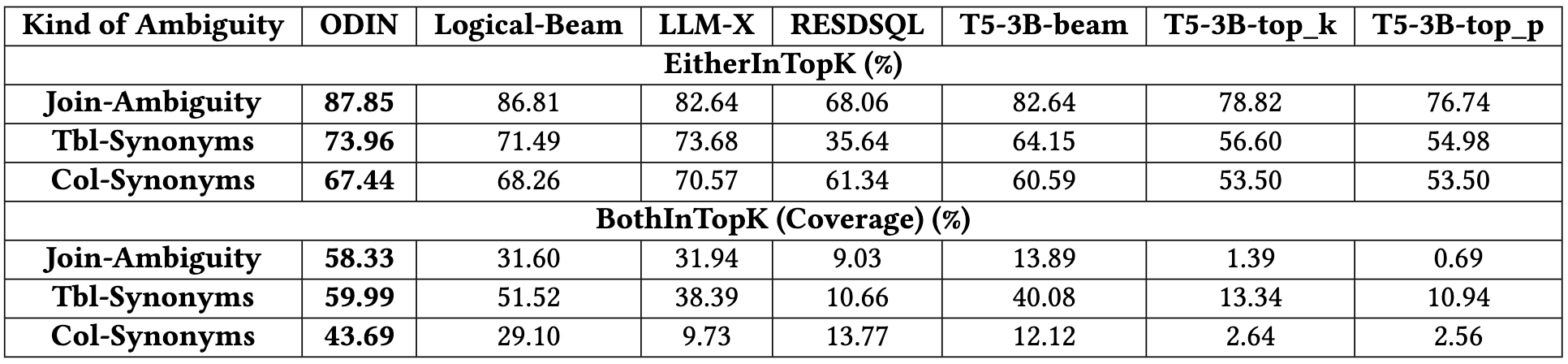}
    \end{center}
    \caption{\namenospace's Generator results alongside various SQL generation baselines for ambiguity types such as Join, Table Synonyms, and Column Synonyms. Each question has two SQL alternatives, and we assess whether either or both are in the generated result set (EitherInTopK and BothInTopK). While most baselines achieve high EitherInTopK accuracies by generating at least one correct alternative, \name outperforms them by generating both alternatives in up to twice as many cases as the next best baseline.}
    \label{fig:expt_2}
\end{figure*}

\begin{figure*}[!h]
\begin{center}
    \includegraphics[width=\linewidth]{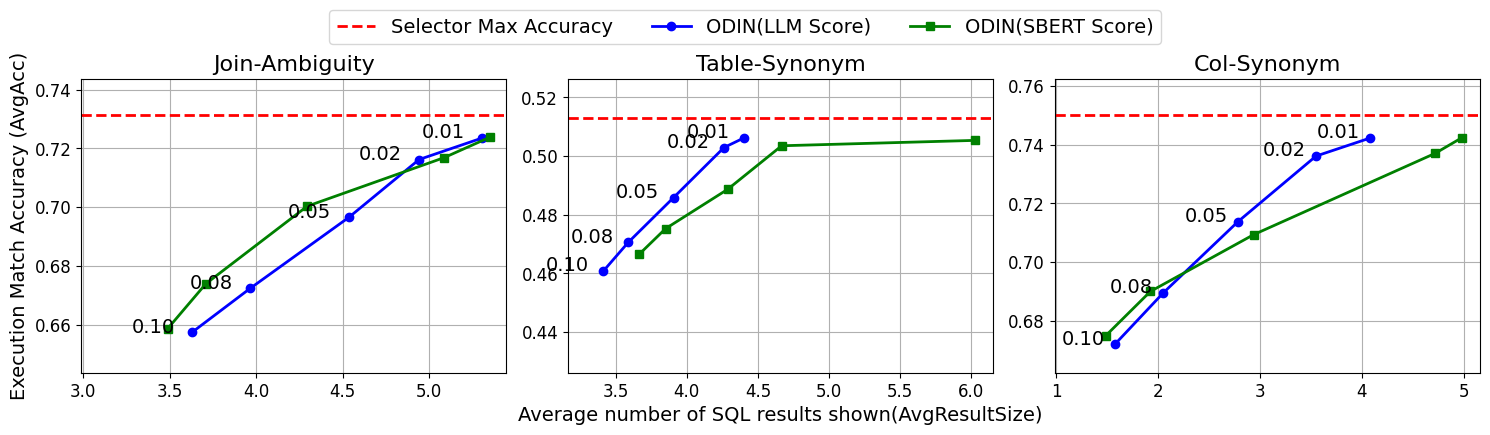}
\end{center}
\caption{\namenospace's Execution Match Accuracy ($AvgAcc$) versus the average number of results shown to the user ($AvgResultSize$) on across three ambiguity types, using various Selector scoring functions (LLM-based and SBERT-based). For each baseline, we vary the Selector's alpha value from 0.01 to 0.1. Higher alpha values lead to more SQL queries being discarded by the Selector, which decreases both accuracy and result set size.}
\label{fig:expt_3}
\end{figure*}

In this experiment, we assess the limitations of traditional diversity-promoting methods in generating ambiguous SQL queries for the NL2SQL task. While these approaches excel at producing a single correct SQL query, subsequent generations often result in minor variations of the same query, missing out on the full spectrum of potential ambiguous interpretations. In contrast, our schema masking-based generation method encourages a broader range of query outputs. To demonstrate this, we conducted evaluations using the AmbiQT benchmark \cite{ambiqt}, which specifically requires generating two distinct and correct SQL alternatives for each input question. The benchmark measures two metrics: \textit{EitherInTopK}, which checks if at least one of the alternatives is present in the generated set, and \textit{BothInTopK} (Coverage), which evaluates whether both alternatives are present. Two SQL queries are deemed equivalent if their execution results match; thus, verifying the existence of a SQL alternative requires matching its execution results with generated SQL queries.

Across all evaluated baselines, the \textit{EitherInTopK} accuracy remains relatively high, indicating that most methods are capable of producing at least one correct query. However, when the requirement shifts to generating both correct alternatives, the accuracy drops significantly. This trend is clearly illustrated in the results shown in \cref{fig:expt_2}. As hypothesized, traditional diversity-promoting techniques excel at identifying one correct SQL query, evident in high \textit{EitherInTopK} scores, but struggle to generate both correct alternatives, leading to lower \textit{BothInTopK} scores.

Our schema masking approach demonstrates a significant advantage, outperforming the next best baseline, Logical Beam, by factors of 1.9$\times$, 1.2$\times$, and 1.5$\times$ in terms of coverage for JOIN ambiguity, Table-Synonyms, and Column-Synonyms, respectively. Additionally, it is noteworthy that different types of ambiguities lead to varying performance across the baselines. LLMs generally handle table-synonyms better than more complex ambiguity types like JOIN ambiguities, as reflected by higher \textit{EitherInTopK} and Coverage scores for table-synonyms. Logical Beam, which employs a specialized decoding strategy to encourage diversity, achieves the second-best performance across most ambiguity types. Nonetheless, the results clearly highlight the superiority of schema masking-based generation, which generates all the ambiguous queries in up to twice as many cases as the next best baseline.

\subsection{Selector Ablation Study}
\label{sec:evaluation:selector}
The two key components of the Selector are the scoring function and the alpha value. In \cref{fig:expt_3}, we illustrate the trade-off points achieved by modifying the scoring function and adjusting the alpha values from 0.01 to 0.1 for each scoring function. For this experiment, we utilize two scoring functions: one based on an LLM and the other based on SBERT. Additionally, we include the maximum possible accuracy of the Selector, represented by the accuracy of \namenospace without the Selector.

\noindent\textbf{Impact of Scoring Function:} An effective scoring function assigns low scores to correct SQL queries and high scores to incorrect ones, improving result pruning. For an alpha of 0.01, the LLM-based scoring displays an average of 5.3, 4.4, and 4.1 SQL queries across three ambiguity classes, while SBERT averages 5.35, 6.1, and 5.9. The LLM-based scoring presents 1.5 to 2 fewer SQL queries than SBERT, although the difference is less pronounced in the case of join ambiguity. While we expect the LLM to outperform SBERT due to its larger model size, SBERT performs better in scenarios where names of ambiguous tables and columns are nearly identical. For example, in a join ambiguity involving the \texttt{birthplace} column in the \texttt{students} table and the similarly named \texttt{birthplace} column in the \texttt{student\_birthplace}, SBERT assigns a similarity score close to 1.0, leading to higher scores for such ambiguous queries. However, SBERT struggles with cases involving synonyms, where the names are less similar.

\noindent\textbf{Impact of Alpha:} The alpha value represents the maximum allowable reduction in accuracy when pruning SQL results from the generation stage. Larger alpha values enable the Selector to prune more SQL queries, resulting in a smaller result set but lower accuracy. Across all three ambiguity classes, increasing the alpha value correlates with a decrease in both accuracy and the number of SQL queries displayed. Higher alpha values allow for a greater drop in recall, leading to stricter thresholds and fewer SQL queries shown to the user. For an alpha value of 0.1, the LLM-based scoring achieves accuracies of 65.7\%, 46.1\%, and 67.2\%, while maximum accuracies are 73.2\%, 51.3\%, and 75.0\%, respectively. Notably, the accuracy with an alpha of 0.1 is approximately 10\% lower than that without the Selector, reflecting the trade-off that the Selector is designed to manage.

\section{Conclusion}
In this paper, we introduced \namenospace, an NL2SQL system that handles schema ambiguity. \name further personalizes SQL results based on user feedback. \name demonstrates consistent accuracy improvements across different ambiguity types in our benchmarks.

\bibliographystyle{ACM-Reference-Format}
\bibliography{text2sql}

\end{document}